Γ-valley assisted intervalley scattering on monolayer and bilayer WS$_2$ revealed by time-resolved Kerr rotation spectroscopy

Huimin Su[1], Aiying Deng[1], Zhiheng Zhen[1], Jun-Feng Dai[1*]

1 Department of Physics, Southern University of Science and Technology, Shenzhen 518055, China

*E-mail: daijf@sustc.edu.cn

Abstract: We investigated the valley depolarization and carrier relaxation process of monolayer and bilayer WS$_2$ at 10 K by using time-resolved Kerr rotation (TRKR) and differential reflectance measurement simultaneously. Two decay processes extracted from TRKR signals were observed on both monolayer and bilayer WS$_2$. In monolayer WS$_2$, the initial ultrafast decay component (< 1 ps) was interpreted as the stimulated emission or Pauli-blocking of electrons and holes in one valley by comparing with carrier decay process. The relatively slow component was around 4 ps under low excitation energy (< 2.21 eV) and then increases with excitation energies, which approaches a saturation value of 15 ps. The onset excitation energy of 2.21 eV suggests the Γ–valley assisted intervalley scattering between the K and K' valley play a critical part during the valley depolarization process under the off-resonance excitation condition. By contrast, the slow decay component (48±1 ps) of bilayer WS$_2$ is comparable with the carrier lifetime (58±0.6 ps). It is attributed to irreversible scattering processes from K (or K') to Γ valley due to its characteristic of indirect band gap semiconductor.

Monolayer transition metal dichalcogenides (TMDCs) are considered as a kind of ideal materials to achieve valley-based optoelectronics [1]. Due to its inversion symmetry broken and strong spin-orbit coupling, energy-degenerate valleys (K and K') at the corners of the Brillouin zone are composed of opposite spin state and coupled with left-handed and right-handed circularly polarized light, respectively. Because of large separation on moment space, this degree of freedom theoretically should be immune to impurities, defects or phonons scattering [2,3]. However, many studies based on various optical measurement techniques, including polarization-resolved photoluminescence [4], time-resolved photoluminescence [5], ultrafast transient absorption spectroscopy [6] and time-resolved Kerr rotation (TRKR) spectroscopy [7,8] have revealed various mechanisms for valley depolarization. Within several picoseconds, strong Coulomb exchange interaction between the K and K' valleys is considered as main scattering mechanism on all the members of TMDCs [6,9]. Moreover, ultrafast defect-assisted recombination [10], ultrafast exciton emission [11] also occur within this time scale. According to $P = \frac{P_0}{1+\tau/\tau_K}$ [5] (P: valley polarization, $P_0$: the theoretical limit of PL polarization, $\tau$: carrier lifetime, $\tau_K$: polarization decay time), these ultrafast processes enlarge the polarization of steady-state photoluminescence by reducing the carrier lifetime. But they do not contribute to intervalley scattering between K and K' valley. In the range of several tens of picoseconds, several valley depolarization origins including the trion depolarization [12,13], defect-related located states [10] and carrier-phonon scattering [14] were studied on monolayer TMDCs. But these processes also do not influence intervalley scattering of carriers. Theoretically, opposite spin orientations in K and K' valleys, protected by time-reversal symmetry, requires a simultaneous spin flip during intervalley scattering [15]. Therefore, besides the origins mentioned above, spin-degenerate Γ-valley is considered as a potential transfer station to achieve carrier scattering between K and K' valley [16,17]. Although some studies have employed this mechanism to discuss their experimental results [1,6], there was no direct experimental evidence.

For TMDCs materials, WM$_2$ (M=S or Se) exhibits larger energy splitting of valence band near K (or K') point than that of MoM$_2$, due to strong spin-orbit coupling in the d-orbitals of the W atom [18]. It is around 400 meV for WM$_2$ based on some experimental reports [18]. For monolayer WSe$_2$, first-principle energy band structure calculations [18-20] show that the top of valence band at Γ-valley is 400 meV lower than that at K-valley, namely $E_v(k) - E_v(\Gamma) = 400\ meV$, which is comparable with spin-splitting energy of valence band at K valley. It indicates a low probability of intervalley scattering through Γ-valley for A exciton at low temperature. Furthermore, fast decay processes through defect-related located states and trions further blur this Γ-valley assisted scattering process for monolayer WSe$_2$ [13]. However, for monolayer WS$_2$, the maximum of valence band at Γ-valley is between spin-split exciton ground-states at K (or K') valley, making intervalley scattering through spin-degenerate Γ-valley feasible in the presence of hot carriers. In this work, we used TRKR spectroscopy and optical differential reflectance spectroscopy to explore a new intervalley scattering channel between K and K' valley. At low temperature (T=10 K), we observed the strong intervalley scattering process within several picoseconds for monolayer WS$_2$, induced by strong Coulomb exchange interaction. This process is accompanied by the renormalization of the quasi-particle band gap. An excitation-dependent decay component was observed on monolayer WS$_2$ under excitation energy higher than 2.21 eV. It is ascribed to the valley depolarization mechanism due to the Γ–valley assisted scattering between K and K' valley. For bilayer WS$_2$, an even slow decay component, which is comparable with carrier lifetime, is induced by irreversible scattering from K (or K') to Γ valley.

The samples studied were monolayer and bilayer WS$_2$ on Si substrate with a 300 nm SiO$_2$ capping layer (Fig. S1(a) and (b) of Supplementary Information (SI)), which were mechanically exfoliated from WS$_2$ bulk single crystal (purchased from 2D semiconductor Inc.). The thickness of monolayer and bilayer samples was confirmed by atomic force microscopy (AFM) and photoluminescence (PL) spectra, as shown in Fig. S1(a), (b) and (c) of SI. The second harmonic generation (SHG) spectra (as shown in Fig. S2(b)), which are resonantly enhanced for interband transition [21], clearly revealed the emission peak of A exciton 1S state at 2.10 eV and B exciton 1S state at 2.45 eV on monolayer WS$_2$. Therefore, the energy splitting of valence band is about 350 meV for monolayer WS$_2$ at T=10 K. The schematic diagram of experimental setup for TRKR measurements is shown in Fig. 1(a). In our experiments, samples were excited by pump pulse with a pulse width of 150 fs and a repetition rate of 80 MHz, generated from a tunable frequency-doubled optical parametric oscillator (OPO) pumped by a 800 nm Ti: sapphire laser. The photon energy of pump pulse was flexibly tuned from 2.14 eV to 2.38 eV, which is between the emission energy of A and B exciton. This can avoid resonant absorption by their ground-state. The helicity of pump beam, e.g. circularly and linearly polarized light, was accurately tuned by controlling the angle between the fast axis of quarter-wave plate and the polarization direction of pump beam. By using different bandpass filters with 10 nm bandwidth, single-frequency probe beam with a pulse width of around 200 ps was selected from a white-light supercontinuum pulse (NKT Photonics Inc.) pumped by the same 800 nm Ti: sapphire laser. Both pump and probe beam were collimated by a non-polarized beam splitter and focused on sample by a 50x objective, with spot sizes of around 4 $\mu m$ and 2 $\mu m$, respectively. During measurements, samples were mounted on a cold-finger of a helium-flow magneto-optical cryostat, which was keep at around 10 K. The reflected probe beam passed through a half-wave plate and a Wollaston prism, then was focused on two diodes of a balanced photodiode bridge (Thorlabs, PDB 210A). By accurately tuning the fast axis of the half-wave plate, we could balance the intensity on two photodiodes to reduce background noise. Moreover, by changing the time delay between two pulses using a mechanical delay stage, the time-dependent intensity difference of two detectors was recorded by a lock-in amplifier, which is proportional to Kerr rotation angle. Meanwhile, the total intensity by summing signals of two photodiodes, which corresponds to carrier lifetime, was recorded by another lock-in amplifier simultaneously. Through this measurement method, the relative time zero between valley and carrier dynamics can be exactly determined.

As shown in Fig. S2(a) of SI, polarization-resolved photoluminescence measurement shows near 24% valley polarization on most representative samples of monolayer $WS_2$ at T=10 K, excited by left-handed circularly polarized ($\sigma^-$) light with energy of 2.09 eV under near-resonant condition. However, Fig. S2(b) of SI shows that for two-photon interband transition with excitation energy from 1.8 eV to 2.8 eV, two right-handed circularly ($\sigma^+$) photons can be simultaneously absorbed at K valley, then generate a $\sigma^-$ SHG signal with near-unity polarization (as shown in the inset of Fig. S2(b) of SI). The corresponding optical selection rule for one- and two- photon interband transitions near K and K' valley is shown in Fig. 1(b). Since SHG process occurs before intravalley or intervalley scattering of carriers, the near-unity polarization of SHG signal illustrates that crystal symmetries related valley-contrasting physics is robust during initial absorption processes around K (or K') valley [22], no matter whether it is resonant absorption or not [23]. It also justifies that strong disorder is absent on our samples, which will weaken valley-dependent selection rule when the excitation is away from the bandgap edge [8]. It is reasonable to conclude that low polarization of steady state photoluminescence is mainly induced by strong carriers scattering between K and K' valley, which is attributed to various physical origins mentioned above.

A linearly polarized probe beam, consisting of a coherent superposition of two opposite-helicity circularly polarized lights, will be absorbed by two valleys at the same time. For the near-resonant absorption, the absorption intensity of two circular components will be different in the presence of an imbalance of carrier population between two valleys, leading to variation of polarized direction of the probe beam reflected from sample. This Kerr rotation angle can be expressed using the following equation:

$$\theta_K \propto N_k - N_{K'} \propto Im[n_+ - n_-] \qquad (1)$$

Where $N_k$ and $N_{k'}$ are carrier density at K and K' point, respectively, $Im[n_+ - n_-]$ is the difference between the image part of complex optical response of two circularly polarized lights.

The representative TRKR and differential reflectance spectra for monolayer $WS_2$ at T=10 K are shown in Fig. 2(a) in the condition of low energy excitation. The energy of pump pulse was fixed at 2.14 eV with bandwidth of 0.04 eV. Compared with emission energy of A exciton at 2 eV on monolayer $WS_2$ shown in Fig. S1(c) of SI, an excess energy of pump pulse is around 140 meV.

While the linearly probe pulse with energy of 2 eV and bandwidth of 0.03 eV, which is nearly resonant with A exciton (2 eV), was used to detect carrier distribution around the band edge of K and K' valley. The purple square line in Fig. 2(a) represents differential reflectance as a function of delay time, which was obtained by summing the intensity of two photodiodes. It corresponds to the carrier lifetime near the band edge. As shown in Fig. S3, there is no visible difference on carrier decay processes under the various excited conditions, including a left-handed circular, right-handed circular and linear excitation, respectively. Since we monitored the total carrier decay processes on two valleys, the carrier scatterings between them do not affect carrier lifetime observed here. For the carrier dynamics, a maximum appears at first 1.6 ps, which corresponds to reflectance enhancement of the probe beam. Then carrier density drops to a minimum at around 10 ps. Subsequently, they return to ground states after 100 ps. Initially, electron-hole pairs are created at the excitation photon energy, then scattering processes with phonons and other carriers lead to carrier population to relax towards the maximum (or minimum) of the valence band (or conduction band) within several hundred femtoseconds [24]. For the reflectance enhancement of probe beam within 1-2ps, we found that it will happen in the presence of stimulated emission process, which was satisfied in the experimental condition of our study [6]. Moreover, the Pauli-blocking of the electron and hole states also prevent the probe beam to be further absorbed, leading to enhancement of reflectance. Subsequently, the renormalization of the quasi-particle band gap due to large population of electron-hole pairs on 2-dimensional materials will lead to the band gap to shrink and the corresponding red shift of optical transition. Therefore, it contributes to the reduced reflectance of probe beam with 10 ps [25]. Finally, electron-hole pairs recombine via radiative or non-radiative processes with decay time of around 40 ps. This recombination time is consistent with the PL lifetime measurement reported in Ref. [26] [26]. Meanwhile, the red, black and blue dots line in Fig. 2(a) show the TRKR spectra under various helicities of pump beam. The power of excitation pulse was lower than 400 μW. In this region, the decay time of normalized TRKR signal keeps unchanged as shown in Fig. S4 of SI. In the case of left-handed circular excitation (red dots line in Fig. 2(a)), two decay processes are adequately fitted by using a bi-exponential function, with the decay time of 0.9 ps and 3.9±0.1 ps, respectively. When the helicity of pump pulse is switched to right-handed circularly polarized light as shown by the black dots line in Fig. 2(a), the sign of Kerr rotation signal reverses simultaneously, as expected from the optical selection rules of intervalley transition on monolayer TMDCs. For linearly polarized excitation, the Kerr rotation signal is negligible due to manual scattering between K and K' valley with the same probability, as shown by the blue dots line in Fig. 2(a). Since Kerr rotation signal is proportional to carrier difference between two valleys, an obvious change of carrier density in one valley will also induce a Kerr rotation signal. Moreover, Reflectance enhancement on carrier lifetime spectrum also happens within this time range. Hence, we believe that the fast decay process (0.9 ps) measured on TRKR data mainly comes from stimulated emission or Pauli-blocking induced reflectance enhancement. The relative slow decay component (around 4 ps) are attributed to strong long-range Coulomb exchange interaction between K and K' valley [9], which mixes the exciton spin states located in different valleys. It has been reported on other members of monolayer TMDCs with characteristic time within several picoseconds [12]. A comparison between valley and carrier dynamics shows that the main scattering process between K and K' valley occurs during band gap renormalization. After that, the concentration of carriers in the two valleys approaches equilibrium. This intervalley scattering process induces rather low valley polarization, which is about 24% circular polarization on steady-state PL measurement.

Next, we performed the TRKR measurement on monolayer $WS_2$ at T=10 K, by tuning the pump energy to 2.38 eV with excess energy of 380 meV, while keeping the energy of probe beam unchanged. Compared with experimental condition mentioned above, this is corresponding to high energy excitation. As shown in the purple square line of Fig. 2(b), the similar carrier decay processes are observed in this case. The enhanced reflectance within 2 ps also supports the ultrafast thermalization and relaxation time (< several hundred femtoseconds) of carrier population in $WS_2$. The Kerr rotation signal (red, black and blue dots line in Fig. 2(b)) exhibits similar two decay processes and helicity-dependence. However, the significant difference is that the slow decay component survives until 35 ps, indicated by dashed rectangle. Moreover, a comparison between Kerr rotation and differential reflectance signal shows that this slow decay process occurs during recombination of electron-hole pairs.

To check evolution of this slow decay component in monolayer $WS_2$ at low temperature, we performed TRKR measurements at various excitation energies, limiting its value among the emission energy of A and B exciton. Fig. 3(a) displays the Kerr rotation angle as a function of delay time with increasing the excitation energy from 2.14 eV to 2.38 eV. The valley lifetime extracted by bi-exponential fitting for various excitations was listed at Table S1 of SI and drawn with error bar in Fig. 3(b). For the photon energy ranging from 2.14 to 2.21 eV, the slow decay time of several picoseconds is independent on the excitation energy. However, when the excitation energy is tuned above 2.21 eV, it gradually increases with increasing the excitation energy. Then it approaches a constant value of around 15 ps for pump energy larger than 2.34 eV. This kind of slow process has been reported on other members of TMDCs family, which was attributed to defect-related localized states on monolayer $WSe_2$, and trion depolarization on monolayer $MoS_2$ [12]. For defect-mediated scattering process, rapid scattering to defect states of excited electron-hole pairs from single valley will contribute to TRKR signal, despite no carrier scattering between K and K' valley. However, we compared typical PL spectra of monolayer $WS_2$ with that of $WSe_2$ reported by previous studies [13]. Some PL peaks below emission of A exciton or trion, which is associated with PL from defect-related localized states reported on $WSe_2$, is absent for monolayer $WS_2$ materials in our study. For trion depolarization process, the long trion emission decay within several tens of picoseconds [13], maybe contribute to slow decay in TRKR signal. However, as shown in inset of Fig. S1(c), the PL spectrum is well fitted with single Gaussian line shape for monolayer $WS_2$ at T=10 K. It supports that trion emission is very weak for the samples in our study. By using 45 meV of electron-exciton binding energy reported in ref. [26] [26], we also performed TRKR measurement with the energy of probe beam of 2.03 eV and 1.97 eV indicated by the arrows in inset of Fig. S1(c), corresponding to a near-resonant absorption with exciton and trion, respectively. As shown in Fig. S5 of SI, the two decay times are 0.8 ps and 4.1±0.5 ps extracted from TRKR spectrum. There is no visible difference on extracted decay time between two energies of probe pulse, indicating negligible influence of trion depolarization on monolayer $WS_2$. Moreover, the timescale for trion decay is within several tens picoseconds [13], which is longer than that in our study. On the other hand, scattering processes including defect-related located states and trion depolarization is independent of excitation energy, as reported on monolayer $WSe_2$ and $MoS_2$ [12], which is completely different from the experimental results in our study. In addition, scattering of bright exciton to an optically forbidden dark state also occurs in this time region, which is accompanied by a decrease of PL intensity with decreasing temperature as reported on monolayer WSe$_2$ [27]. However, for monolayer WS$_2$, our temperature-dependent PL measurements (Fig. S1(d)) reveal opposite trend to reported results. So this mechanism can also be safely excluded.

According to first-principle energy band structure calculations[18-20], the energy difference between maximum of Γ-valley and K-valley is around 200 meV. In our measurement, the difference between the onset energy of slow decay component (2.21 eV) and emission energy of A exciton (2 eV) is around 0.21 eV, which is consistent with the calculation results. Thus this excitation-dependent phenomenon can be better understood by considering Γ-valley assisted scattering processes between K and K' valley. Electron-hole pairs are firstly created on the excited states on the K valley, and then relaxation to the K valley extremum or scattering to the Γ-valley occurs on high excitation energy. It is followed by the scattering of carriers in the Γ-valley with spin degeneracy return to the K and K' valley with different scattering possibilities. Such process modifies the valley population contrast and hence contributes to the valley polarization. The schematic diagram of Γ-valley assisted intervalley scattering between K and K' valley is shown in Fig. 3(c).

Now let us consider the exciton rate equation when Γ-valley assisted intervalley scattering channel exists. After high energy states on K valley are pumped by a circularly polarized light, the evolution of the exciton populations ($N_K$ and $N_{K'}$ on K and K' valleys) near the band edges can be described by the following rate equations as

$$\frac{dN_K}{dt} = \frac{N_K^*}{\tau^*} - \frac{N_K}{\tau} + \frac{N_\Gamma}{\tau_{\Gamma K}} - \frac{N_K}{\tau_v} + \frac{N_{K'}}{\tau_v} \tag{2}$$

$$\frac{dN_{K'}}{dt} = -\frac{N_{K'}}{\tau} + \frac{N_\Gamma}{\tau_{\Gamma K'}} + \frac{N_K}{\tau_v} - \frac{N_{K'}}{\tau_v} \tag{3}$$

where $N_K^*$ and $\tau^*$ is the population and the intravalley relaxation time constant of photon-generated exciton on the excited states. $\tau_v$ is the intervalley scattering time constant between the K and K' valley due to strong Coulomb exchange interaction. $N_\Gamma$ is the hole population on the Γ-valley. And $\tau_{\Gamma K}^{-1}$ and $\tau_{\Gamma K'}^{-1}$ is the scattering rate of holes from the Γ-valley to the K and K' valley, respectively. $N_K^*$ decays quickly within the initial several 100s of femtoseconds after the pump pulse, so we ignore its influence for intervalley scattering. Hence the rate equations of the valley-dependent scattering process with (4-1) or without (4-2) the intermediate Γ-valley are:

$$\frac{d\Delta N_K}{dt} \approx -\frac{\Delta N_K}{\tau} - 2\frac{\Delta N_K}{\tau_v} \tag{4-1}$$

$$\frac{d\Delta N_K}{dt} \approx -\frac{\Delta N_K}{\tau} - 2\frac{\Delta N_K}{\tau_v} + N_\Gamma(\frac{1}{\tau_{\Gamma K}} - \frac{1}{\tau_{\Gamma K'}}) \tag{4-2}$$

where $\Delta N_K = N_K - N_{K'}$, corresponding to our Kerr rotation signal. In the presence of Γ-valley assisted intervalley scattering, the different scattering rates between Γ → K and Γ → K' valley also affects the valley depolarization process. In our case, we have $\tau_{\Gamma K}^{-1} \neq \tau_{\Gamma K'}^{-1}$, due to the inequivalent bandgap renormalization at the K and K' valley. This process leads to a modification on the valley depolarization curve and slow down the rapid decay of the Kerr rotation signal during the first several tens of picoseconds, which is consistent with the relaxation time of the resonance energy shift of the A exciton observed in Ref. [25] [25].

We also employed TRKR spectroscopy to study the valley-dependent scattering process on bilayer $WS_2$, which exhibits near 80% valley polarization at 10 K under near-resonant excitation measured by polarization-resolved photoluminescence spectrum as shown in Fig. S6. This robust valley polarization on bilayer $WS_2$ has also been reported on Ref. [26] [28] [26,28]. Fig. 4(a) shows valley and carrier dynamics for bilayer $WS_2$ at 10 K, with the energy of pump and probe beam at 2.14 eV and 2 eV, respectively. By comparing with peak position (2.03 eV) of PL spectrum of bilayer $WS_2$, the energy of probe beam is also near resonant with A exciton. As shown in the purple line of Fig. 4(a), the carrier lifetime shows a single exponential decay with characteristic time of 58 ps for bilayer $WS_2$, which is longer than that of monolayer one (40 ps). It is mainly because of its indirect nature of the electronic band gap (Fig. S1(c)), with the maximum of valence band at Γ-point for bilayer $WS_2$. Therefore, this decay component corresponds to relaxation process of energetic electron-hole pairs on K valley to Γ-valley by interacting with phonons. For Kerr rotation spectra shown in red, black and blue line of Fig. 4(a), it exhibits the same helicity-dependence as that on monolayer one, which means that the left-handed (right-handed) circularly polarized light generate positive (negative) Kerr signal and linearly polarized light gives negligible one. By fitting with bi-exponential function, the decay process of Kerr rotation signal on bilayer $WS_2$ can be exactly extracted, with decay time of 0.43 ps and 48±1 ps, respectively. Due to the reduced binding energy of exciton and Coulomb screening effect [29], exchange interaction is much weaker on bilayer sample than that on monolayer one, leading to faded scattering between K and K` valley. Around 80% circular polarization on bilayer measured by polarization-resolved PL also supports this point of view. Hence, we believe that the fast decay component (0.43 ps) are not induced by Coulomb exchange interaction between two valleys. Further studies are needed to clarify this process. Moreover, in contrast with monolayer $WS_2$, the slow decay component (48±1 ps) is comparable with carrier lifetime (58±0.6 ps) on bilayer $WS_2$, as indicated by the inset of Fig. 4(a). This slow decay component on Kerr signal can be well explained in the presence of the item of $\frac{\Delta N_K}{\tau}$ ($\tau$ : carrier lifetime) in equation (4-1). Due to weak intervalley scattering, we have $\Delta N_K = N_k - N_{k'} \approx N_k$. Therefore, the carrier recombination process contributes to slow decay component in Kerr rotation signal. Since there is a large energy difference between Γ and K point (around 250 meV) [1], it is impossible for carriers to scatter back to K or K' valley on bilayer $WS_2$ at low temperature[1], as diagramed in Fig. 4(b). Combining the mechanism of weak intervalley scattering through Coulomb exchange interaction and Γ-valley assisted intervalley scattering, high valley polarization is maintained on bilayer $WS_2$.

In conclusion, we use TRKR and differential reflectance measurement to detect the intervalley scattering channel on monolayer and bilayer $WS_2$ at low temperature. Our results reveal that the ultrafast valley dynamics (< 5 ps), induced by strong Coulomb exchange interaction, is the main mechanism of valley depolarization on monolayer $WS_2$ under low excitation energy. Moreover, when the excitation energy is increased, an even slow decay component observed on TRKR signal is turned on, which is attributed to Γ-valley assisted carrier scattering between K and K' valley for monolayer $WS_2$. For bilayer one, the slow decay component is also observed on TRKR signal, but it is comparable with carrier lifetime. Due to its characteristic of indirect band gap semiconductor, it is attributed to carrier scattering process from K to Γ valley. The scattering channels through Coulomb exchange interaction and Γ-valley assisted intervalley are restricted on bilayer $WS_2$. Therefore, high valley polarization is maintained as observed on polarization-resolved PL under near-resonant excitation.

We would like to thank Prof. Xiao-Dong Cui, and Prof. Hai-Zhou Lu for helpful discussions. This work is supported by the National Natural Science Foundation of China (11204184, 11604139), and Special Funds for the Development of Strategic Emerging Industries in Shenzhen (no. JCYJ20150630145302235).

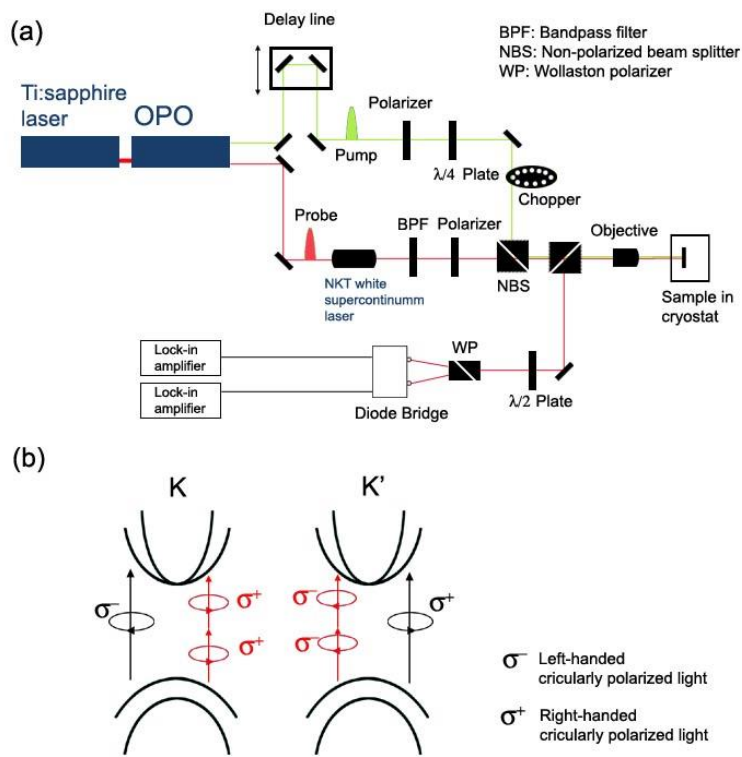

Figure 1. (a) A schematic diagram of TRKR setup. (b) Diagram of valley-dependent optical selection rule for one- (black arrows) and two-photon (red arrows) interband transitions near K and K' valley.

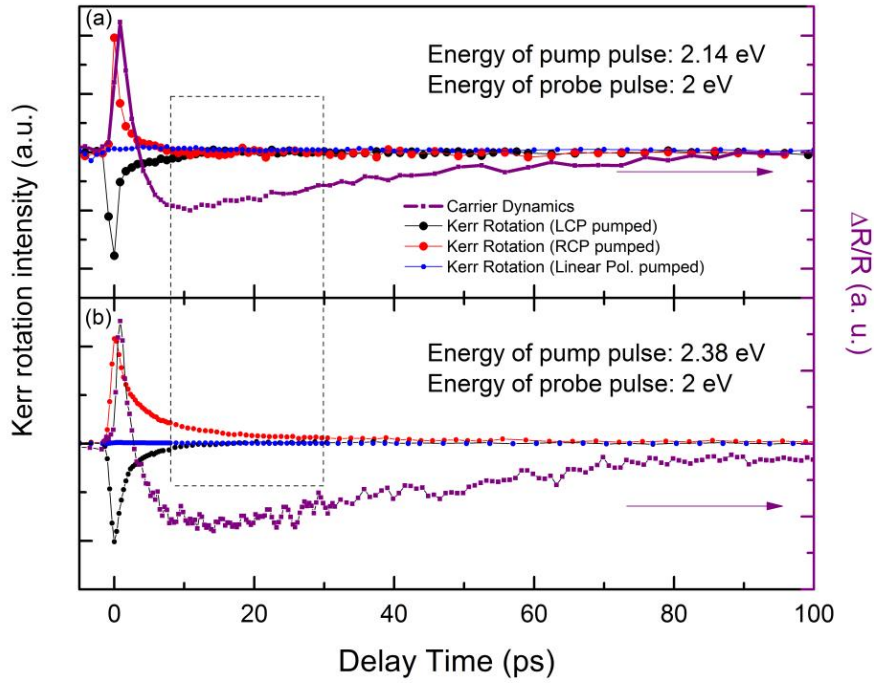

Figure 2. (a) and (b) Valley and carrier dynamics simultaneously measured by TRKR experiment at T=10 K, with the excitation energy of 2.14 eV and 2.38 eV, respectively. The energy of probe pulse is fixed at 2 eV, corresponding to near-resonant excitation of A exciton on monolayer $WS_2$. An even slow decay component is observed in the case of high excitation energy indicated by dashed square. $\frac{\Delta R}{R} = \frac{R - R_{pump}}{R}$, where R and $R_{pump}$ denote the reflectance of the $WS_2$ sample without and with pump beam.

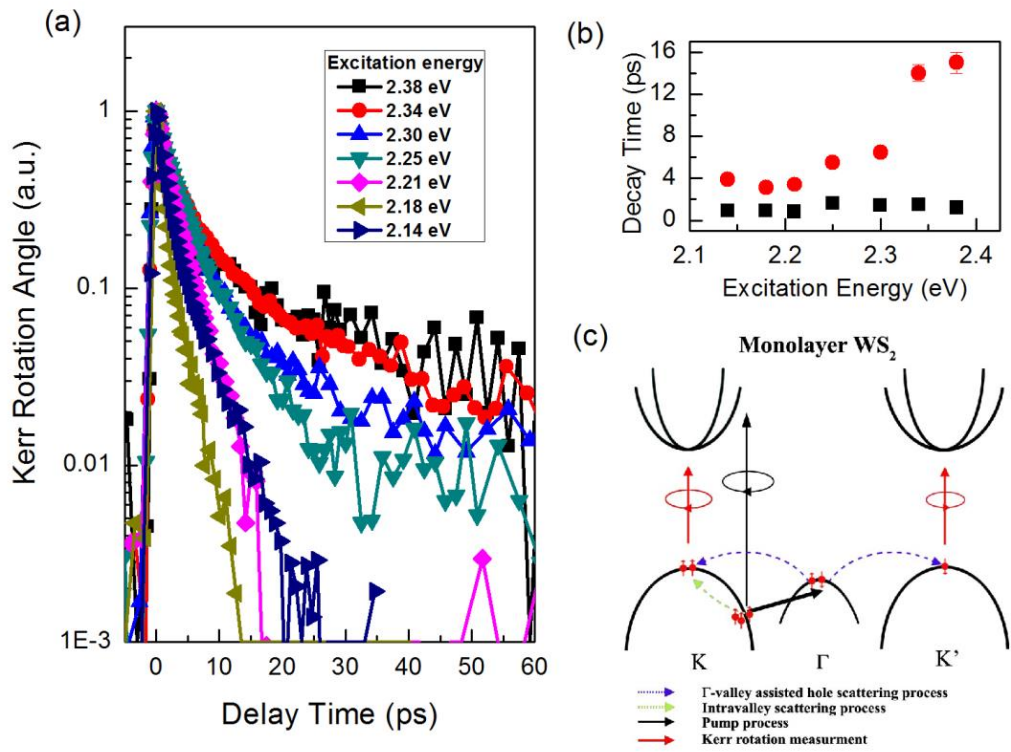

Figure 3. (a) The Kerr rotation angle as a function of delay time for excitation energy ranging from 2.14 eV to 2.38 eV. (b) The fast and slow decay time extracted from TRKR spectra (c) A schematic diagram of Γ-valley assisted intervalley scattering between K and K' valley.

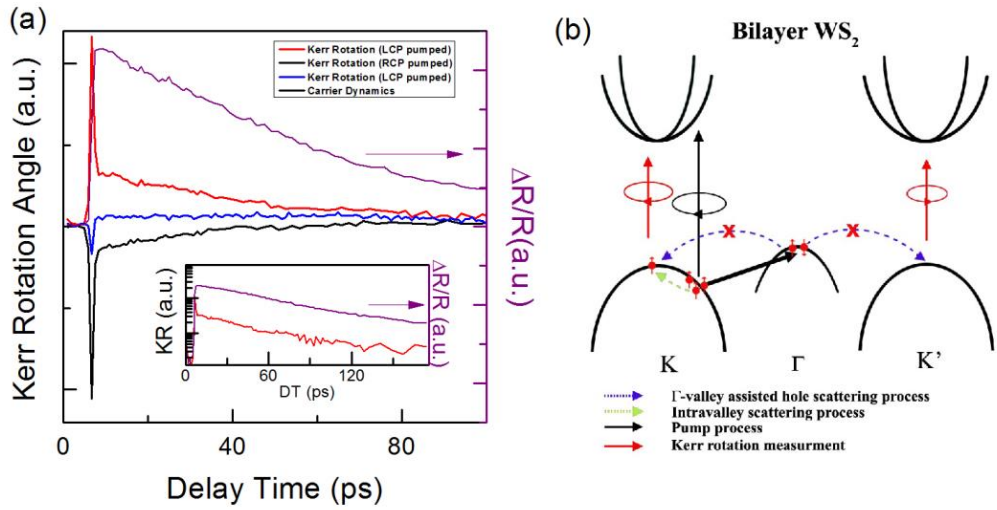

Figure 4. (a) Valley and carrier dynamics simultaneously measured by TRKR at T=10 K for bilayer $WS_2$, with the energy of pump and probe beam at 2.14 eV and 2 eV, respectively. (b) A schematic diagram of carrier scattering process between Γ-valley and K (or K') valley for bilayer $WS_2$.